
%
\font\eightrm=cmr8
\font\eighti=cmmi8
\font\eightsy=cmsy8
\font\eightbf=cmbx8
\font\eighttt=cmtt8
\font\eightit=cmti8
\font\eightsl=cmsl8
\font\sixrm=cmr6
\font\sixi=cmmi6
\font\sixsy=cmsy6
\font\sixbf=cmbx6
\catcode`@11
\newskip\ttglue
\font\grrm=cmbx10 scaled 1200

\def\eightpoint{\def\rm{\fam0\eightrm}
\textfont0=\eightrm \scriptfont0=\sixrm \scriptscriptfont0=\fiverm
\textfont1=\eighti \scriptfont1=\sixi \scriptscriptfont1=\fivei
\textfont2=\eightsy \scriptfont2=\sixsy \scriptscriptfont2=\fivesy
\textfont3=\tenex \scriptfont3=\tenex \scriptscriptfont3=\tenex
\textfont\itfam=\eightit \def\it{\fam\itfam\eightit}
\textfont\slfam=\eightsl \def\sl{\fam\slfam\eightsl}
\textfont\ttfam=\eighttt \def\tt{\fam\ttfam\eighttt}
\textfont\bffam=\eightbf
\scriptfont\bffam=\sixbf
\scriptscriptfont\bffam=\fivebf \def\bf{\fam\bffam\eightbf}
\tt \ttglue=.5em plus.25em minus.15em
\normalbaselineskip=6pt
\setbox\strutbox=\hbox{\vrule height7pt width0pt depth2pt}
\let\sc=\sixrm \let\big=\eightbig \normalbaselines\rm}
\newinsert\footins
\def\newfoot#1{\let\@sf\empty
  \ifhmode\edef\@sf{\spacefactor\the\spacefactor}\fi
  #1\@sf\vfootnote{#1}}
\def\vfootnote#1{\insert\footins\bgroup\eightpoint
  \interlinepenalty\interfootnotelinepenalty
  \splittopskip\ht\strutbox 
  \splitmaxdepth\dp\strutbox \floatingpenalty\@MM
  \leftskip\z@skip \rightskip\z@skip
  \textindent{#1}\footstrut\futurelet\next\fo@t}
\def\fo@t{\ifcat\bgroup\noexpand\next \let\next\f@@t
  \else\let\next\f@t\fi \next}
\def\f@@t{\bgroup\aftergroup\@foot\let\next}
\def\f@t#1{#1\@foot}
\def\@foot{\strut\egroup}
\def\footstrut{\vbox to\splittopskip{}}
\skip\footins=\bigskipamount 
\count\footins=1000 
\dimen\footins=8in 

\def\ref#1{$^{#1}$}
\def\flex{\raise 6pt\hbox{$\leftrightarrow $}\! \! \! \! \! \! }
\def\oversome#1{ \raise 8pt\hbox{$\scriptscriptstyle #1$}\! \! \! \! \! \! }
\def\tr{ \mathop{\rm tr}}

\newbox\bigstrutbox
\setbox\bigstrutbox=\hbox{\vrule height10pt depth5pt width0pt}
\def\bigstrut{\relax\ifmmode\copy\bigstrutbox\else\unhcopy\bigstrutbox\fi}
\def\refer[#1/#2]{ \item{#1} {{#2}} }
\def\rev<#1/#2/#3/#4>{{\it #1\/} {\bf#2}, {#3}({#4})}
\def\boxit#1{\vbox{\hrule\hbox{\vrule\kern3pt
\vbox{\kern3pt#1\kern3pt}\kern3pt\vrule}\hrule}}

\def\sqr#1#2{{\vcenter{\hrule height.#2pt
   \hbox{\vrule width.#2pt height#1pt \kern#1pt
    \vrule width.#2pt}
    \hrule height.#2pt}}}

\def \E {{{\rm e}}}
\def \tr {{\rm tr}\, }

\magnification 1200
\nopagenumbers
\hfill CERN-TH/95-81

\hfill hep-th/9503235
\vskip 1cm
\centerline {\grrm  Loop scattering in two-dimensional QCD}
\vskip .6cm

\centerline {E. Abdalla$^{(1),(2)}$\newfoot {${}^*$}{Permanent address:
Instituto de F\'\i sica - USP, C.P. 20516, S. Paulo, Brazil.} and M.C.B.
Abdalla$^{(1)}$\newfoot{${}^\dagger $}{Permanent address: Instituto de
F\'\i sica Te\'orica - UNESP, R. Pamplona 145, 01405-000, S. Paulo, Brazil.}}
\vskip .4cm

\centerline{(1) CERN-TH}

\centerline{ CH-1211 Geneva 23 - Switzerland}

\vskip .2cm

\centerline {(2) Institut f\"ur Theoretische Physik, Universit\"at Heidelberg}

\centerline { Philosophenweg 16, 6900 Heidelberg 1- Germany}
\vskip 3cm
\centerline{\bf Abstract}
\vskip .5cm

\noindent Using the integrability conditions that we recently obtained in
QCD$_2$ with massless fermions, we arrive at a sufficient number of
conservation laws to be able to fix the scattering amplitudes involving a
local version of the Wilson loop operator.

\vfill
\noindent CERN-TH/95-81
\vskip .3cm
\noindent hep-th/9503235
\vskip .3cm
\noindent March 1995
\eject
\countdef\pageno=0 \pageno=1
\newtoks\footline \footline={\hss\tenrm\folio\hss}
\def\folio{\ifnum\pageno<0 \romannumeral-\pageno \else\number\pageno \fi}
\def\advancepageno{\ifnum\pageno<0 \global\advance\pageno by -1
\else\global\advance\pageno by 1 \fi}

Two-dimensional QCD has been studied by several authors (see ref. [1] for a
review). The theory presents, in a clear way, several features expected to
characterize strong interactions, but besides such desirable physical features,
it is possible to understand several of its properties analytically. In fact,
we claim that it may be possible to obtain its full on-shell solution, as far
as the scattering of closed loop (white) operators is concerned.

We represent the fermionic determinant in terms of a bosonic functional
integral over the (exponential of $i$ times the) WZW action,\ref{2,3}
and use the Polyakov--Wiegmann identity\ref{2} to split the action of
the product of two fields in terms of the WZW action of each single field
and a contact term; this shows that, at the Lagrangian level, the dynamics
factorizes in terms of several conformally invariant theories (WZW and
ghost actions), and a non-conformally invariant piece, given by a WZW action
perturbed off criticality, which contains the (main) dynamical information.
Finally, these fields, decoupled at the Lagrangian level, are coupled via BRST
constraints\ref{4} obeyed by the theory, implying that only white objects
appear in the physical subspace. The most important result implying
the possibility of computing the $S$-matrix is the integrability of the
afore-mentioned off-critically perturbed WZW theory.\ref{5}

Let us briefly summarize the known results. The two-dimensional QCD partition
function with massless quarks is given by
$$
\eqalignno{
{\cal Z}\left[i_\mu,\eta,\overline\eta\right]= &\int{\cal D}\psi{\cal D}
\overline\psi{\cal D}A_\mu{\cal D}E{\cal D}[{\rm ghosts}]\cr
& \times \E^{iS_{{\rm ghosts}}+i\int{\rm d}^2x\left[\overline\psi i\!\not
D\psi-
{1\over 2}\tr E^2+{1\over 2}\tr EF_{+-} + i^\mu A_\mu + \overline\eta\psi+
\overline\psi\eta\right]}\quad .&(1)\cr}
$$

The fermion may be integrated out in terms of the gauge potentials  $U$ and
$V$,
introduced as
$$
A_+ = {i\over e}U^{-1}\partial _+U \quad , \quad
A_- = {i\over e}V\partial _-V^{-1} \quad ; \eqno(2)
$$
these in turn, are implemented in the quantum case via inclusion of the
corresponding Jacobian in (1). As a result one obtains the WZW action
upon
use of
$$\displaylines{\det i\!\!\not\!\!D=\E^{i\Gamma[UV]} \cr
\noalign{\hbox{and}}
\hfill \det\left({\partial A_+\over\partial U} {\partial A_-\over\partial V}
\right)=\det i\!\!\not\!\!D^{{\rm adj}}=\E^{ic_V\Gamma[UV]}\; ,\hfill (3)\cr}$$
where a gauge-invariant regularization prescription has been chosen to evaluate
the Jacobian, fixing the contact term $A_+A_-$; therefore, the result is a
function of the gauge-invariant product $UV$; the well-known WZW functional
reads
$$
\Gamma [U]=   {1\over 8\pi}\int {\rm d}^2x\, \partial ^\mu U^{-1}\partial _\mu
U +{1\over 4\pi}\int_0^1{\rm d}r \int {\rm d}^2x\, \epsilon^{\mu\nu} \widetilde
U^{-1}\dot {\widetilde U} \widetilde U^{-1}\partial_\mu \widetilde U
\widetilde U^{-1} \partial _\nu \widetilde U ,\eqno(4a)
$$
obeying the Polyakov--Wiegmann identity
$$
\Gamma[UV]=\Gamma  [U]+\Gamma[V]+{1\over 4\pi}\tr\int{\rm d}^2x\,
U^{-1}\partial_+UV\partial_-V^{-1} \quad .\eqno(4b)
$$
In terms of the gauge-invariant field $\Sigma = UV$, the auxiliary $E$-field
interaction becomes, upon introduction of $F_{+-}$ as a function of $U$ and
$V$,
$$
\tr EF_{+-}={i\over e}\tr UEU^{-1}\partial_+\left(\Sigma\partial_-\Sigma^{-1}
\right)={i\over e}\tr\widetilde E\partial_+\left(\Sigma\partial_-\Sigma^{-1}
\right)\quad .\eqno(5)
$$

We change the auxiliary variable as
$$\partial_+\widetilde E={ie(c_V+1)\over 2\pi}\beta^{-1}\partial_+
\beta\; ,\eqno(6)$$
and use again the Polyakov--Wiegmann identity. This procedure  leads to a
complete separation of variables at  the Lagrangian level, as we see from
$$
\eqalignno{
&{\cal Z}\left[i_\mu,\eta,\overline\eta\right]=\int{\cal D}\widetilde g
\,\E^{i\Gamma[\tilde g]}{\cal D}U{\cal D}\left[{\rm ghosts}\right]\,
\E^{iS_{{\rm ghosts}}}
\int {\cal D}\widetilde\Sigma\,\E^{-i(c_V+1)\Gamma[\tilde\Sigma]}\cr
&\times \int {\cal D} \beta\, \E^{i\Gamma[\beta] + {\mu^2i\over 2}\tr \int
{\rm d}^2 z\,[\partial_+^{-1}(\beta^{-1}\partial_+\beta)]^2}\,\E^{i\int{\rm
d}^2
z\,i_\mu A_\mu -i\int{\rm d}^2z\,{\rm d}^2w\overline\eta(z)(i \not D)^{-1}(z,w)
\eta (w) }\quad ,&(7)\cr}
$$
where $\tilde g=UgV$, $\widetilde\Sigma =\beta\Sigma$ and $\mu=(c_V+1)e/2\pi$.
The quadratic and non-local piece, which we symbolize by $\Delta$, can be
linearized and made local by the introduction of the identity
$$
\E^{{i\over 2}\mu^2\Delta}=\int{\cal D}C_-\,\E^{i\int{\rm d}^2x\,{1\over 2}\tr
(\partial_+C_-)^2-\mu\tr\int{\rm d}^2x\,C_-(\beta^{-1}\partial_+\beta)}\quad ,
\eqno(8)
$$
where $C_-$ plays the role of an auxiliary field. The theory contains hidden
constraints, as discussed elsewhere.\ref{4,5} Such constraints are essential
to build the asymptotic states of the theory, which cannot have colour.

The non-trivial dynamical content of the theory is described by the $\beta$
field, which is the only non-conformally invariant piece of the theory. It is
useful to discuss the model in parallel to its dual, obtained by rewriting
the field $C_-$ appearing  in eq. (8) as $C_-= {i\over 4\pi\mu}W
\partial_-W^{-1}$. After a simple change of variables  and use of (4$b$),
one obtains for the $\beta\, ,\, C_-$ integration the following partition
function (below, we use $\widetilde \beta = \beta W$):
$$
{\cal Z} = \int {\cal D} \tilde \beta \,\E^{i\Gamma[\widetilde \beta]}\int
{\cal D} W\,\E^{-i(c_V+1)\Gamma[W]-{i\over 2(4\pi\mu)^2} \tr\int {\rm d}^2 z
\, [\partial_+(W\partial_- W^{-1})]^2}\quad ,\eqno(9a)
$$
or equivalently, the dual action
$$
S= -(c_V+1) \Gamma[W] + \tr {1\over 2}\int{\rm d}^2x\,\left(-B^2+{1\over 2\pi
 \mu } \partial _+ B \partial _-WW^{-1}\right)\quad ,\eqno(9b)
$$
where the field $B$ plays, in the dual case, the same role as $C_-$ did before,
namely it is an auxiliary field.
The theory (7) and its dual (9) can be studied by simple methods. The
$\beta$ (resp. $W$) equations of motion correspond to a conservation law
$$\displaylines{
\hfill\partial_+\left(I_-^{\beta ,W}\right) =0\; ,\hfill (10)\cr
\noalign{\hbox{where}}\cr
\hfill I_-^\beta (x^-) = 4\pi \mu^2 J_-^\beta (x^+,x^-) + \partial_+ \partial_-
J_-^\beta (x^+,x^-) + [J_-^\beta (x^+,x^-), \partial_+ J_-^\beta(x^+,x^-)]
\hfill (10a)\cr
\hfill I_-^W (x^-)={(c_V+1)\over 4\pi}J_-^W+{1\over  (4\pi\mu)^2}
\partial _+\partial_-J_-^W+{1\over (4\pi\mu)^2}[J_-^W,\partial J_-
^W] \; . \hfill (10b)\cr}$$

Notice that for the current $I_-^W(x^-)$ we have only local expressions.
Concerning the current $I_-^\beta$, the first class constraints are given by
$$
\eqalignno{
i\widetilde g\partial_-\widetilde g^{-1}-i(c_V+1)\Sigma\partial_-\Sigma^{-1}
+J_-({\rm ghosts})& \sim 0\quad ,&(11a)\cr
i\widetilde g^{-1}\partial_+\widetilde g-i(c_V+1)\Sigma^{-1}\partial_+\Sigma
+J_+({\rm ghosts})& \sim 0\quad ,&(11b)\cr}
$$
leading to two BRST nilpotent charges $Q^{(\pm)}$, as discussed in
refs. [4, 5]. Therefore we find two first-class constraints. For the dual
theory one finds a very similar structure (see [1, 5]). Second class
constraints also show up, being given by
$$
\displaylines{\hfill \Omega_{ij}^\beta=(\beta \partial _-\beta^{-1})_{ij} +
4i\pi \mu  (\beta C_-\beta
^{-1})_{ij} - (\widetilde g \partial _-\widetilde g^{-1})_{ij}\quad ,\hfill
(12a) \cr
\hfill\Omega_{ij}^W= (c_V+1) \Sigma^{-1}\partial _+ \Sigma - (c_V + 1)
W^{-1}\partial
_+W + {1\over \mu} W^{-1} \partial _+ B W =0 \quad . \hfill (12b)\cr}$$

To obtain the consequences of such a huge set of conservation laws and
constraints, we have to verify how the corresponding charges act on asymptotic
states. This action can be unravelled once one knows the Lorentz transformation
properties of the charges. A useful set of conserved charges is
$$
Q^{(n)}= \int {\rm d}x\, (t-x)^n I_-(x)\quad . \eqno(13)
$$
It is not difficult to obtain
$$
[T, Q^{(n)}] = n\, Q^{(n)}\quad ,\eqno(14)
$$
where the Lorentz generator $T$ acts on an  asymptotic one-particle state as a
derivative with respect to the rapidity $\theta$:
$$
T = {d\over d\theta }\quad ,\eqno(15)
$$
from which one obtains
$$
Q^{(n)} \simeq (p_-)^n J \quad , \eqno(16)
$$
where $J$ is the generator of the left transformations for the $\beta$ fields.
This would mean that the first charge, $Q^{(1)}$, is the generator of
right-$SU(N)$ transformations for the $\beta$ fields! In the quantum theory
there is no contribution from the short-distance expansion of $J_-$ and
$\partial_\mu J_-$, since divergences are too mild. The $SU(N)$ transformation
generators are simple to compute. In the $\beta$ language one has, for
left-$SU(N)$ transformations:
$$
\eqalign{
J^{L_{SU(N)}}_+&=0\cr
J^{L_{SU(N)}}_-&=-{1\over 2\pi}\partial_-\beta\beta^{-1}+i\mu\beta
C_-\beta^{-1}\cr}\quad ,\eqno(17a)
$$
while for right-$SU(N)$ transformations we obtain
$$\eqalign{
J^{ R_{SU(N)}}_-&=i\mu C_-+\left[ \partial_+C_-,C_-\right]\cr
J^{ R_{SU(N)}}_+&=-{1\over 2\pi}\beta^{-1}\partial_+\beta\cr}
\quad . \eqno(17b)
$$
For the first set, i.e. the left transformations, one finds that the
currents are equivalent to analogous chiral currents corresponding to
free fields. On the other hand, the right transformations lead to an
infinite number of
conservation laws due to the presence of the Lax pair, and  reproduce
eq. (10); therefore we have confirmed the fact that (10) generates the
right-$SU(N)$ transformations. For the left transformations, we obtain only
left-moving currents, which explain  constraints such as (12), where they
are expressed in terms of free WZW currents.

The structure is even clearer in the $W$ language, where the right-$SU(N)$
transformations are
$$
\eqalign{
J^{R_{SU(N)}}_+&=0\quad ,\cr
J^{R_{SU(N)}}_-&=-{c_V+1\over 2\pi}W^{-1}\partial_+W+{1\over 4\pi\mu}
W^{-1}\partial_+BW\quad ,\cr}\eqno(18)
$$
while the left-$SU(N)$ transformations are
$$
\eqalign{
J^{ L_{SU(N)}}_-&={c_V+1\over 4\pi} \partial_-WW^{-1} +{1\over
4\pi\mu}\left[ B, \partial_-WW^{-1}\right]\quad ,\cr
J^{ L_{SU(N)}}_+&=-{1\over 4\pi\mu}\partial_+B\quad .\cr}\eqno(19)
$$

In the $\beta_{ai}$ language the left indices, such as $a,b,c,d$ are thus free,
described by a trivial $S$-matrix, while the right indices, such as $i,j,k,l$
are described by an $SU(N)$ covariant integrable $S$-matrix, which has a
well-known classification.\ref{6} We thus write the (unique) $\beta$-ansatz as
$$
\eqalignno{
\langle ai\theta_1, bj\theta_2\vert ck\theta_3dl\theta_4\rangle =\, &\delta(
\theta_1-\theta_3)\delta(\theta_2-\theta_4)\delta_{ac}\delta_{bd}(\sigma_1(
\theta)\delta_{ik}\delta_{jl}+\sigma_2(\theta)\delta_{il}\delta_{jk})\cr
& + \delta (\theta_1-\theta_4)\delta(\theta_2-\theta_3)\delta_{ad}\delta_{bc}
(\sigma_1(\theta)\delta_{il}\delta_{jk}+\sigma_2(\theta)\delta_{ik}\delta_{jl}
)\quad ,&(20)\cr}
$$
where $\theta=\theta_1-\theta_2$ is the relative rapidity. A similar result
for the dual theory exists, where the left/right definition has to be
interchanged, namely the field is $W_{ia}$ in the dual case.

There are four classes of possible (non-trivial) $S$-matrices. Since the
$\beta$
field represents a kind of Wilson-loop-type variable as given by (5), we
suppose that, for large $N$, $\beta^2\sim \beta$, which implies an $S$-matrix
of class II as classified in ref. [6], or $\sigma_2= {2\pi i\over N}\sigma_1$,
and a bound state
$$
B_{N=\infty}(\theta)={{\rm sh}(\theta)+i{\rm sin}(\pi/3)\over
{\rm sh}(\theta) -  i{\rm sin}(\pi/3)}\quad .\eqno(21)
$$

Finally, the only asymptotic states compatible with gauge $SU(N)$ colour
symmetry, are those confining such degrees of freedom; therefore we have the
scattering elements\ref{7}
$$
\langle \theta_1\theta_2\vert \theta_3\theta_4\rangle = B_N
(\theta)\left[\delta(\theta_1-\theta_3)\delta(\theta_2-\theta_4) +
\delta(\theta_1-\theta_4)\delta(\theta_2-\theta_3)\right]\quad ,\eqno(22)
$$
where, for large $N$, the amplitude is given by (21).
\vskip 1cm
\noindent {\bf Acknowledgements}

\noindent This work was partially supported by CAPES Brazil, by the
World Laboratory, and the Alexander von Humboldt Stiftung.
\vskip 1cm
\penalty-300
\centerline {\bf References}
\vskip 1cm
\nobreak
\refer[[1]/E. Abdalla and M.C.B. Abdalla, preprint CERN-TH/95-49,
hep-th/9503002, to appear in Phys. Rep.]

\refer[[2]/A.M. Polyakov and P.B. Wiegmann, Phys. Lett. {\bf B 131} (1983)
121; {\bf B 141} (1984) 223]

\refer[[3]/E. Witten, Commun. Math. Phys. {\bf 92} (1984) 455]

\refer[[4]/D. Karabali and H.J. Schnitzer, Nucl. Phys. {\bf B 329} (1990)
649]

\refer[[5]/E. Abdalla and M.C.B. Abdalla, Phys. Lett. {\bf  B 337} (1994)
347--353; and  preprint CERN-TH/7354-94, hep-th/9407128, to appear in Int. J.
Mod. Phys. {\bf A}]

\refer[[6]/B. Berg, M. Karowski, V. Kurak, P. Weisz, Nucl. Phys. {\bf B 134}
(1978) 125.]

\refer[[7]/E. Abdalla, M.C.B. Abdalla and K. Rothe, {\it Non-perturbative
methods in two-\-dimen\-sional quantum field theory} (World Scientific,
Singapore, 1991)]
\end